\documentclass[final]{svjour3}
\usepackage[dvipdfmx]{graphicx}
\usepackage{rotating}
\usepackage{amssymb}
\usepackage{mathptmx}
\makeatletter
\journalname{Journal of Low Temperature Physics}

\providecommand{\eprint}[2][]{\url{#2}}
\usepackage[sort&compress,numbers]{natbib}
\usepackage{doi}
\usepackage{url}
\usepackage{hyperref}

\usepackage{comment}
\usepackage{xcolor,colortbl}
\usepackage{graphicx}
\usepackage{adjustbox}
\usepackage[misc]{ifsym}

\begin{document}

\newcommand{\hdblarrow}{H\makebox[0.9ex][l]{$\downdownarrows$}-}

\title{Monte-Carlo Simulations of Superconducting Tunnel Junction Quantum Sensors for the BeEST Experiment}

\author{C.E. Bray$^1\!\!$\and L.J. Hiller$^2\!\!$\and K.G. Leach$^1\!\!$\and S. Friedrich$^2\!\!\!\!\!$} 

\institute{\Letter~~ Connor E. Bray \\ $~\quad~~$ cbray@mines.edu \\[5pt] $^1\quad$ Colorado School of Mines, Golden, CO 80401, U.S.A\\[3pt] $^2\quad$ Lawrence Livermore National Lab, Livermore, CA 94550, U.S.A.}

\maketitle

\begin{abstract}

Superconducting Tunnel Junctions (STJs) are used as high-resolution quantum sensors to search for evidence of sterile neutrinos in the electron capture decay of $^7$Be. We are developing spatially-resolved Monte-Carlo simulations of the energy relaxation in superconductors to understand electron escape after the $^7$Be decay and distinguish details in the STJ response function from a possible sterile neutrino signal. Simulations of the charge generation and the Fano factor for different materials agree with the literature values. Initial simulations of the escape tail are consistent with observations, and contain fine structure in the line shape. The line shape will be refined as better experimental data become available.


\end{abstract}

\section{Introduction}

The Beryllium Electron capture in Superconducting Tunnel junctions (BeEST) experiment uses momentum reconstruction of nuclear Electron Capture (EC) decay in $^7$Be to perform a model-independent search for the existence of heavy neutrino mass states \cite{StephanPRL}. For EC decay, the neutrino mass can be accessed by measuring the kinetic energy of the daughter nucleus,

\begin{equation}
    \label{eqn:recoilDaughter}
    T_D = \frac{Q_{EC}^2-m_\nu^2c^4}{2(Q_{EC}+m_Dc^2)}.
\end{equation}

\begin{figure}[ht]
    \centering
    \includegraphics[width=0.45\textwidth]{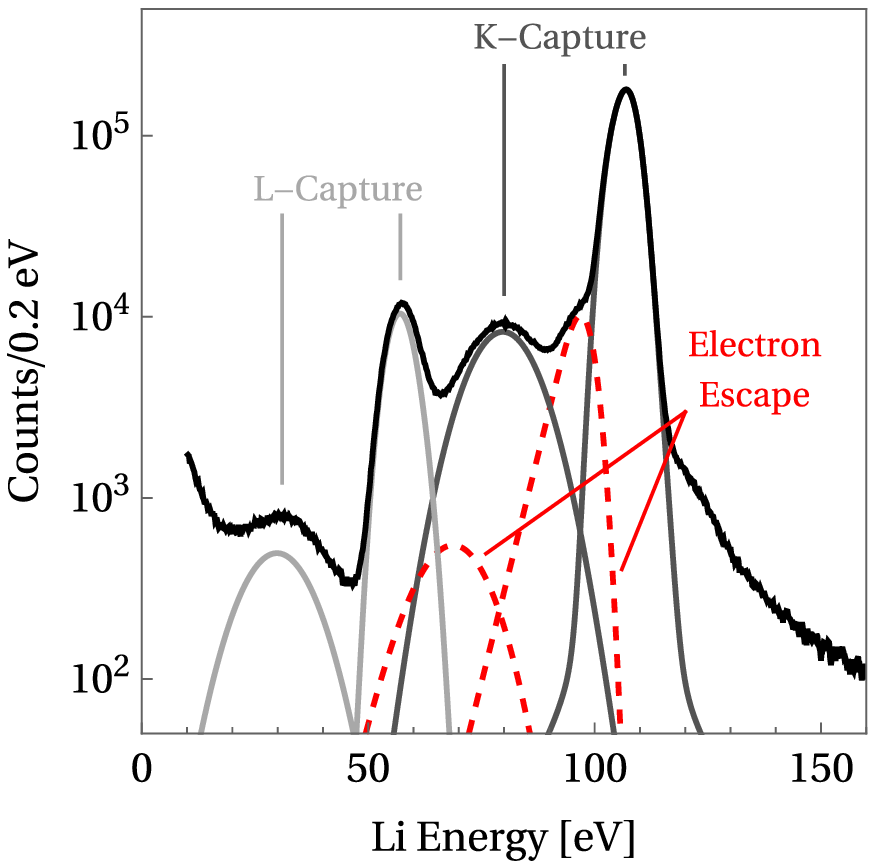}
    \hspace{0.02\textwidth}
    \includegraphics[width=0.45\linewidth]{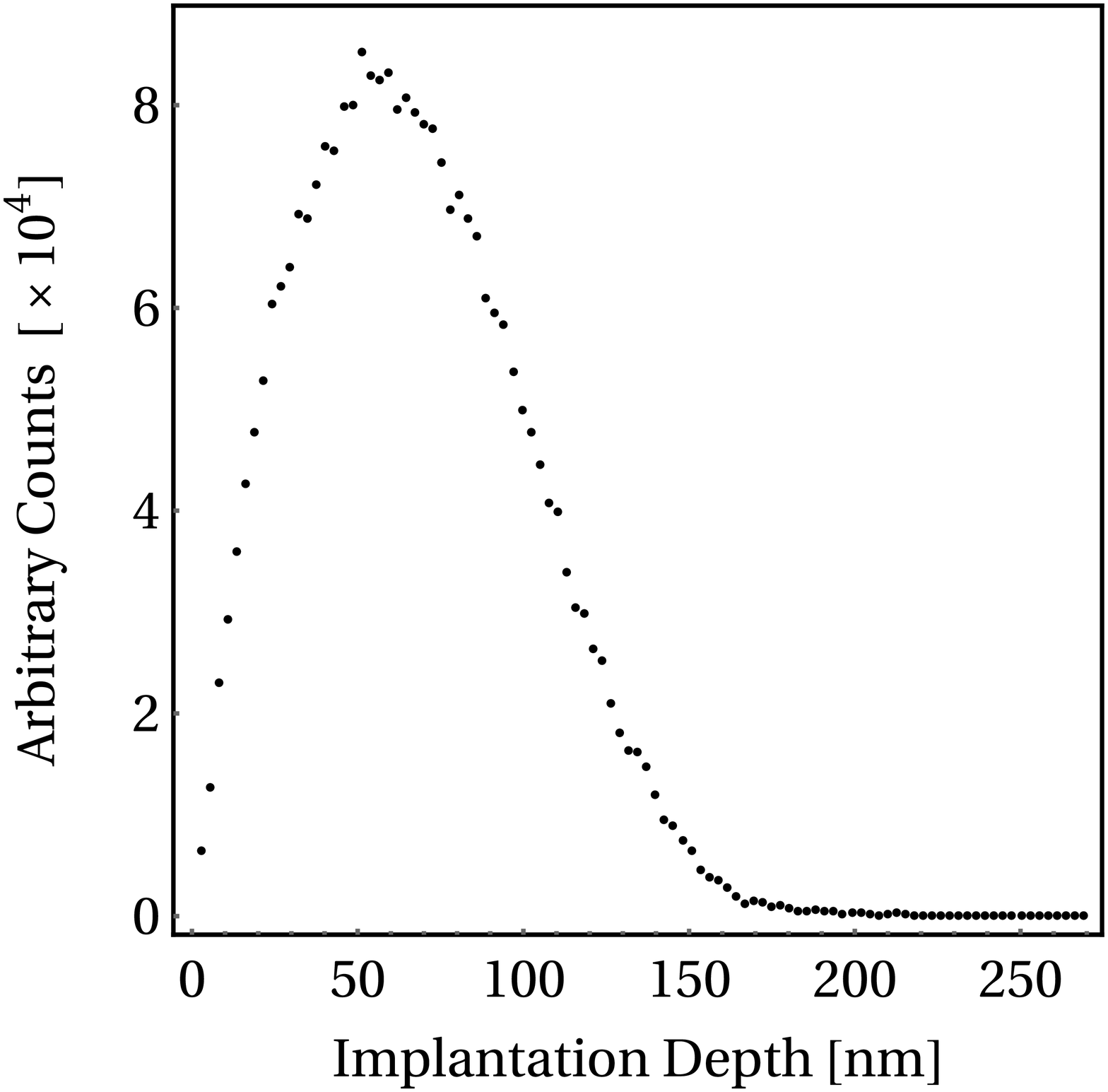}
    \caption{The measured $^7$Li recoil spectrum (black) shows four peaks for the different $^7$Be decay channels. Fits to the four decay channels and to the observed electron escape tails from the K-Capture peaks are shown (color online). Other fit features not shown. Figure adapted from \cite{StephanPRL} \label{fig:experimentalSpectrum}\vspace*{-6px}}
    \caption{SRIM simulated ion depth distribution from a 25 keV ion implantation\cite{ZIEGLER20101818} at TRIUMF-ISAC \label{fig:SRIM}}
    
\end{figure}

Here $Q_{EC}$ is the energy released in the EC decay, $m_\nu$ is the neutrino mass, and $m_D$ is the mass of the daughter atom. The experimental signal shown in Figure \ref{fig:experimentalSpectrum} from the $^7$Be sample has four peaks predicted by the Standard Model (SM) which correspond to the probabilities of a K-shell or L-shell electron being captured, and the probability of decaying into the ground state of $^7$Li or to an excited state. A heavy neutrino signal would appear as an offset spectrum from the active neutrino background at some lower energy and intensity, depending on the mass and mixing angle of the sterile neutrino, respectively. 

To capture the entire recoil energy, the radioactive $^7$Be is implanted directly into an STJ sensor at 25 keV by the TRIUMF Isotope Separator and Accelerator (ISAC) rare-isotope beam facility in Vancouver, Canada \cite{Dilling2014}. The simulated distribution of ion implantation depths generated by a Stopping and Range of Ions in Matter (SRIM) simulation\cite{ZIEGLER20101818} (Figure \ref{fig:SRIM}) shows that a significant number of the $10^8$ implanted ions reside within only a few nm of the surface of the detector. From those ions, it is expected that the 56 eV Auger electron generated after a K-capture decay could have sufficient energy to escape through the surface of the detector and cause the low-energy tails that we observe in the experimental spectrum in Figure \ref{fig:experimentalSpectrum}. The tails currently limit the sensitivity in the BeEST experiment for low neutrino mass. We are developing a spatially-resolved Monte-Carlo code to characterize these tails and understand the underlying effects.

\section{Spatially-resolved Monte-Carlo Simulations of the Energy Relaxation}

Our Monte-Carlo simulation of the energy relaxation cascade is based on the Drude model and tracks individual quasiparticles and phonons through the entire process. Each particle travels along a straight line until it interacts, and the two reaction products are emitted in random but opposite directions while conserving energy. Throughout the simulation, the interaction length is determined by an exponential distribution with a mean set by the particle’s mean free path at its current energy.

The simulation starts with a single electron with an energy $E = 56$ eV corresponding to the Li KLL Auger electron energy. The electron initially loses energy by exciting other electrons as it travels through the material. We assume a mean free path taken from reference \cite{ziaja2006} that has been determined by a fit to measurements of low-energy electron ionization ranges. For materials without published low-energy measurements, the mean free path is scaled to account for the different electron density. Electrons continue to lose energy by interactions with other electrons until the mean electron-electron scattering rate falls below the phonon emission rate and energy relaxation by phonon emission starts to dominate. 

The rate at which an electron with energy E emits a phonon with energy $\Omega$ is determined by the material-dependent electron-phonon coupling strength $\alpha^2$ and the available density of states for phonons $F(\Omega)$ and for electrons $Re\left(\frac{E}{(E^2-\Delta^2)^{1/2}}\right)$. The total phonon emission rate is calculated by integration over all phonon energies according to \cite{kaplanPhysRevB.14.4854}

\begin{equation}
\label{eqn:tauP}
\tau_P^{-1}(E) = \frac{2\pi}{\hbar Z_1(0)} \int_0^{E-\Delta} d\Omega \alpha^2F(\Omega) \textnormal{Re}\left[ \frac{E-\Omega}{\left( (E-\Omega)^2-\Delta^2 \right)^{1/2}} \right] \left( 1- \frac{\Delta^2}{E(E-\Omega)} \right).
\end{equation}

Values of $\alpha^2F(\Omega)$ have been measured for many materials \cite{ATLASkhotkevich_yanson_1995}, and values for the renormalization factor $Z_1(0)$ are taken from \cite{kaplanPhysRevB.14.4854}. The phonon emission rate is converted to a mean free path by assuming all electrons move at the Fermi velocity, and the actual interaction length during the simulations is again determined by an exponential distribution with that mean. The phonon energy $\Omega$ is determined by sampling a distribution given by the integrand of Equation \ref{eqn:tauP}, which ensures the original quasiparticle remains above-gap. To save computation time, the electron-energy-dependent phonon energy distribution is tabulated for each $\alpha^2F(\Omega)$ before the main simulation. This allows the simulation to generate a single random integer and look up the associated phonon energy, linearly interpolating between the table steps of $\Delta/100$. Once the phonon energy $\Omega$ is determined, the electron energy E is reduced by the same amount and the propagation directions for both particles are once again randomized.

If the emitted phonons have energies $\Omega > 2\Delta$, they break Cooper pairs according to the electron-phonon coupling strength $\alpha^2$ and the density of states available for the two quasiparticles. Integration over all final electron energies gives a pair-breaking rate of

\begin{equation}
\label{eqn:tauB}
\tau_B^{-1}(\Omega) = \frac{4\pi N_F \alpha^2(\Omega)}{\hbar I} \int_\Delta^{\Omega-\Delta} \frac{dE}{(E^2-\Delta^2)^{1/2}} \frac{E(\Omega-E)+\Delta^2}{\left( (\Omega-E)^2-\Delta^2 \right)^{1/2}},
\end{equation}

\noindent
where $N_F$ is the electron density of states at the Fermi energy in the normal state and $I$ is the ion density of the material \cite{kaplanPhysRevB.14.4854}. We convert this rate to a mean free path assuming phonon propagation at the speed of sound \cite{steinberg1996equation}. The distribution of quasiparticle energies is determined by sampling the integrand of Equation \ref{eqn:tauB}, and the directions of the resulting quasiparticles are again randomized. Both phonon emission and pair breaking continue until all electrons have relaxed to energies $E < 3\Delta$, and cannot produce an above-gap phonon, and all phonons to energies $\Omega < 2\Delta$ so that the total number of quasiparticles can no longer change. 

Our simulations go beyond earlier Monte-Carlo simulations \cite{KURAKADO1982275,RANDO1992173,hiller_2001}, in that they include the initial phase of the energy relaxation that is dominated by electron-electron interactions and that they track each individual particle and its position and energy. They also differ from earlier spatial simulations \cite{ZehnderPhysRevB.52.12858}, in that they do not assume a local thermal equilibrium and diffusive energy propagation but follow each particle individually. This is made possible by significant advances in computational power since those publications.

The simulations allow implementing a simple model to describe a suspected signal loss mechanism for the BeEST experiment. In this model, any electron that reaches the STJ surface with an energy above the work function leaves the detector and does not contribute to the recorded energy. This is, of course, a simplification of any real surface, where oxides, adsorbates, and imperfections may modify the work function and introduce other loss mechanisms. Nonetheless, it is a starting point to understand the microscopic origin of details in the response function and can be refined as better experimental data become available.

\section{Simulation Results}

We first test our code by reproducing the results of earlier Monte-Carlo simulations for different superconductors. Surface effects are excluded by making the detector infinitely large. For each material, $10^5$ events were run, each starting with a single electron with an energy between 1 and 56 eV. As expected \cite{KURAKADO1982275}, the cascade statistics show no dependence on the initial electron energy in this range. We then calculate the average energy $\epsilon \equiv \frac{Q}{<N>}$ that is required to produce a single excess quasiparticle and the Fano factor $F \equiv \frac{<(N-<N>)^2>}{<N>}$ that quantifies the fluctuations in the number of quasiparticles. The results agree well with the published values, supporting the simplification of earlier simulations to neglect electron-electron scattering. The electron-phonon coupling function is known experimentally for Ta, Al, and Nb \cite{ATLASkhotkevich_yanson_1995}, and we assume the form of the $\alpha^2F(\Omega)$ function to be quadratic for Hf to extend the simulations to that material, which has potential for future STJ detectors due to its small energy gap of 0.021 meV. In all cases, the simulations confirm the earlier results of $\epsilon \approx 1.7\Delta$ and $F \approx 0.2$ (Table \ref{tab:comp}).

\definecolor{Gray}{gray}{0.85}
\definecolor{LightCyan}{rgb}{0.88,1,1}
\definecolor{LightGreen}{rgb}{0.88,1,0.88}

\begin{table}[h]
         \centering
         \caption{Comparison of $\epsilon$ and Fano Factor Calculations for Sn, Nb, Ta, Al, and Hf \cite{KURAKADO1982275,RANDO1992173,hiller_2001} \label{tab:comp}}
\begin{tabular}{r|cll}
Reference               & Material & $\epsilon/\Delta$ & F        \\ \hline
Kurakado\cite{KURAKADO1982275} & Sn       & 1.68              & 0.195(1) \\
Rando\cite{RANDO1992173}       & Nb       & 1.747             & 0.22(1)  \\
Hiller\cite{hiller_2001}     & Ta       & 1.76              & 0.230(5) \\
Hiller\cite{hiller_2001}     & Al       & 1.71              & 0.216(4) \\
Hiller\cite{hiller_2001}     & Nb       & 1.71              & 0.214(2) \\
This Work               & Ta       & 1.78              & 0.233(2) \\
This Work               & Al       & 1.71              & 0.210(5) \\
This Work               & Nb       & 1.72              & 0.214(2)\\
This Work               & Hf       & 1.72              & 0.206(3)
\end{tabular}
\end{table}

\begin{figure}[h]
\centering
\includegraphics[width=\linewidth]{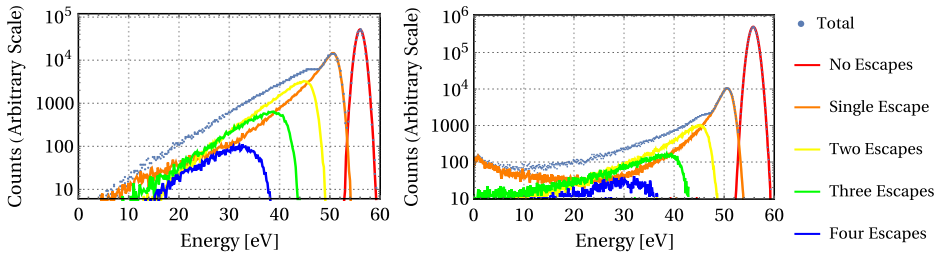}
\caption{Impulse response to a 56 eV Auger electron from a fixed 10 nm implantation depth (left) and from an initial depth distribution derived from the SRIM simulation in Figure \ref{fig:SRIM} (right), broadened to match the detector resolution (color online). Note the deviation from an EMG tail shape caused by 2+ electron escape, particularly at $\sim$10 eV below the primary peak. \label{fig:impulse}}
\end{figure}

To demonstrate the advantages of particle tracking for the BeEST experiment, we simulate the impulse response to electrons from a constant implantation depth in a Ta-based STJ. The electrons have an initial energy of 56 eV corresponding to Li KLL Auger electron energy produced after the K-capture decay of $^7$Be, and they are emitted isotropically. Figure \ref{fig:impulse} (left) shows the response of electrons from an implantation depth of 10 nm, broadened by the STJ detector resolution of 2 eV. The escape tail encompasses 35\% of the total events, and its shape roughly follows an Exponentially-Modified Gaussian (EMG) with a decay scale of $6.33 \pm 0.03$ (stat.) eV. For an implantation depth of 20 nm, the tail contains only 9.0\% of the events and decays with a characteristic energy scale of $2.50 \pm 0.01$ (stat.) eV. In both cases, the tail is offset from the primary peak by the 4.5 eV work function of Ta. Interestingly, the tail shows some fine structure roughly $\sim$10 eV below the primary peak. This fine structure is due to the small integer number of electrons that escape from the STJ, all of which require at least an energy of 4.5 eV and therefore have an energy distribution with a sharp onset. It remains to be seen if this effect is present in actual detectors where different loss mechanisms with different loss and onset energies are likely to be present.

The simulation is repeated for an initial electron depth distribution taken from the $^7$Be implantation into Ta at an energy of 25 keV (Figure \ref{fig:SRIM}). The escape tail is then a convolution of the implantation depth distribution with the depth-varying tail lengths and fractions (Figure \ref{fig:impulse}, right). As a result, it can no longer be fit to a single EMG function. Notably, the fraction of events that escape with a large percentage of the initial electron energy is greater than an EMG tail would predict due to the increased fraction of events $\lesssim10$ nm from the surface of the STJ. Above 20 eV, the tail can be fit with the sum of two EMG functions, with characteristic energy scales of $1.62 \pm 0.01$ (stat.) eV and $8.7 \pm 0.1$ (stat.) eV, centered at 51.4 eV and 46.9 eV respectively. The fraction of events in the tail is found to be 4.9\% of the total, reflecting the importance of events close to the surface.

\section{Discussion}

The simulations suggest that energy deposition close to the detector surface can generate low-energy tails in the response function due to electron escape during the initial relaxation cascade. In some cases, this tail has a roughly exponential shape and can be approximated by an EMG function. This supports earlier analyses where such a shape has been observed \cite{ONeil2020,meganLineShape}. In other cases, details of the escape tail deviate from a simple EMG, especially when the source is distributed over varying depths below the surface. An interesting result is the observation of fine structure in the simulated tails. It is due the small integer number of electrons that escape from the surface and depends on the assumption that there is a single signal loss mechanism with a sharp low-energy cutoff. A single, well-defined work function is unlikely in actual devices due to surface imperfections and oxides, and it remains to be seen if the resulting fine structure will be observed experimentally.

In the BeEST sterile neutrino experiment, the source distribution is given by the implantation profile of $^7$Be in Ta-based STJs at an energy of 25 keV (Figure \ref{fig:SRIM}), and the Li KLL Auger electrons emitted after a K-capture decay of $^7$Be are emitted isotropically. Under these conditions, our simulations predict that the escape tail contains 4.9\% of the total number of events. Given the uncertainties of the mean free path and applicability of the Drude model, we consider this fraction of 4.9\% to be consistent with the observed 6.7\% low-energy tail from the first phase of the BeEST experiment (Figure \ref{fig:experimentalSpectrum}) \cite{SpencerPRL}.

In our simulations, the characteristic energy scale of the more intense single-electron escape tail is $1.62 \pm 0.01$ (stat.) eV. The longer tail with an energy scale of $8.7 \pm 0.1$ (stat.) eV will only be visible in high-statistics low-background spectra. This is lower than the scales in earlier experiments with AuBi-TES and HgTe-Si microcalorimeters, which range from 10 to 25 eV \cite{ONeil2020,meganLineShape}, but in the same range as the 4.3 eV scale observed in the BeEST experiment \cite{SpencerPRL}. This difference likely reflects the lower electron energies involved in the BeEST experiment, which have shorter mean free paths and are therefore more easily absorbed in the detector \cite{ziaja2006}. Future extensions of our simulations will investigate more closely which parameters of the simulations need to be adjusted to match the experimental data more closely.

\section{Conclusions}

We are developing spatially-resolved Monte-Carlo simulations of the energy relaxation cascade in superconducting tunnel junction quantum sensors. They include the first stage of the cascade that is dominated by electron-electron scattering. The simulations reproduce the established values of $\epsilon \approx 1.7\Delta$ for the average energy to generate a quasiparticle and the Fano factor $F \approx 0.2$ for different materials. Our initial simulations assume electron escape at the sensor surface for energies above the work function as the only signal loss mechanism. In some cases, they generate an exponentially decaying tail below the primary peak whose shape and magnitude are consistent with earlier experiments. In others, they predict fine structure in the tail due to the small discrete number of emitted electrons. The simulations will be refined as better experimental data become available to assess whether the fine structure can mimic a signal in the BeEST sterile neutrino search.

\begin{acknowledgements}
This work was supported by the DOE-SC, Office of Nuclear Physics under grant DE-SC0021245 and by the LLNL Laboratory Directed Research and Development program through Grants 19-FS-027 and 20-LW-006. This work was performed under the auspices of the U.S. Department of Energy by Lawrence Livermore National Laboratory under Contract No. DE-AC52- 07NA27344.

Data will be made available upon reasonable request.
\end{acknowledgements}

\bibliography{bib.bib}

\end{document}